\documentclass[11pt]{article}
\usepackage{graphicx}

\newcommand{\BABARPubYear}    {04}

\newcommand{\BABARConfNumber} {047}
\newcommand{\SLACPubNumber} {10619}

\input pubboard/babarsym
 
\def\pipi     {\ensuremath{\pi\pi}}

\def\fish    {\ensuremath{\cal F}}
\def\akpi  {\ensuremath{{\cal A}_{K\pi}}}

\def\fpm {\ensuremath{f_{\pm}(\deltat)}}

\def\spipi {\ensuremath{S_{\pi\pi}}}
\def\cpipi {\ensuremath{C_{\pi\pi}}}
\def\de {\ensuremath{\Delta E}}

\def\Btag {\ensuremath{B_{\rm tag}}}

\def\Bflav {\ensuremath{B_{\rm flav}}}

\long\def\inst#1{\par\nobreak\kern 4pt\nobreak
    {\it #1}\par\vskip 10pt plus 3pt minus 3pt}

\begin{document}
{\pagestyle{empty}

\begin{flushright}
\babar-CONF-\BABARPubYear/\BABARConfNumber \\

SLAC-PUB-\SLACPubNumber \\

August 2004 \\
\end{flushright}

\par\vskip 5cm

% Title of the paper
\begin{center}
{\Large \bf \boldmath 
Improved Measurements of \CP-Violating Asymmetries \\
in $\Bz\to\pip\pim$ Decays}
\end{center}
\bigskip

\begin{center}
\large The \babar\ Collaboration\\
\mbox{ }\\
\today
\end{center}
\bigskip \bigskip

% Abstract
\begin{center}
\large \bf Abstract
\end{center}
We present preliminary updated measurements of the CP-violating parameters 
$\spipi$ and $\cpipi$ in $\Bz\to\pip\pim$ decays. Using a sample of $227$ million 
$\Y4S\to\BB$ decays collected with the \babar\ detector at the \pep2\ asymmetric-energy 
$\epem$ collider at SLAC, we observe $467\pm 33$ signal decays and measure 
$\spipi = -0.30\pm 0.17\,({\rm stat})\pm 0.03\,({\rm syst})$, and 
$\cpipi = -0.09\pm 0.15\,({\rm stat})\pm 0.04\,({\rm syst})$.

\vfill
\begin{center}

Submitted to the 32$^{\rm nd}$ International Conference on High-Energy Physics, ICHEP 04,\\
16 August---22 August 2004, Beijing, China

\end{center}

\vspace{1.0cm}
\begin{center}
{\em Stanford Linear Accelerator Center, Stanford University, 
Stanford, CA 94309} \\ \vspace{0.1cm}\hrule\vspace{0.1cm}
Work supported in part by Department of Energy contract DE-AC03-76SF00515.
\end{center}

\newpage
} % end of pagestyle{empty}

% Input author list file
%
%author list removed temporarily to save trees 7/9/04 RNC
%
\begin{center}
\small

The \babar\ Collaboration,
\bigskip

%% author list as of 02-Jul-2004 (609 authors)
%
B.~Aubert,
R.~Barate,
D.~Boutigny,
F.~Couderc,
J.-M.~Gaillard,
A.~Hicheur,
Y.~Karyotakis,
J.~P.~Lees,
V.~Tisserand,
A.~Zghiche
\inst{Laboratoire de Physique des Particules, F-74941 Annecy-le-Vieux, France }
% ...

\end{center}\newpage

% The body of the paper starts here
\section{INTRODUCTION}
\label{sec:Introduction}
In the standard model, all \CP-violating effects arise from 
a single phase in the Cabibbo-Kobayashi-Maskawa quark-mixing matrix~\cite{CKM}.  
In this context, neutral $B$ decays to the \CP\ eigenstate 
$\pip\pim$ can exhibit mixing-induced \CP\ violation through interference between 
decays with and without $\Bz$--$\Bzb$ mixing, and direct \CP\ violation through 
interference between the $b\to u$ tree and $b\to d$ penguin decay processes~\cite{pipicpv}.  
Both effects are observable in the time evolution of the asymmetry between $\Bz$ and $\Bzb$ 
decays to $\pip\pim$, where mixing-induced \CP\ violation leads to a sine oscillation 
with amplitude $\spipi$ and direct \CP\ violation leads to a cosine oscillation with 
amplitude $\cpipi$.  In the absence of the penguin process, $\cpipi = 0$ and $\spipi = \stwoa$, with
$\alpha \equiv \arg\left[-V_{\rm td}^{}V_{\rm tb}^{*}/V_{\rm ud}^{}V_{\rm ub}^{*}\right]$,
while significant tree-penguin interference leads to $\cpipi \ne 0$ and 
$\spipi = \sqrt{1 - \cpipi^2}\sin{2\alpha_{\rm eff}}$. 
The difference between $\alpha_{\rm eff}$ and $\alpha$ can be determined from a
model-independent analysis using the isospin-related decays $B^{\pm}\to\pipm\piz$ and 
$\Bz,\,\Bzb\to\piz\piz$~\cite{alphafrompenguins,isospin}.

The Belle collaboration recently reported~\cite{BelleSin2alpha2004} 
an observation of \CP\ violation in $\Bz\to\pip\pim$ decays using a data sample
of $152$ million $\BB$ pairs, while our previous measurement~\cite{BaBarSin2alpha2002} 
on a sample of $88$ million $\BB$ pairs was consistent with no \CP\ violation.  In this 
paper we report improved measurements of the \CP-violating parameters $\spipi$ and 
$\cpipi$ using a data sample comprising $227$ million $\BB$ pairs collected with the 
\babar\ detector at the \pep2\ asymmetric-energy $\epem$ collider at SLAC.

\section{THE \babar\ DETECTOR}
\label{sec:babar}
The \babar\ detector is described in detail elsewhere~\cite{ref:babar}.  
The primary components used in this analysis are a 
charged-particle tracking system consisting of a five-layer silicon 
vertex tracker (SVT) and a 40-layer drift chamber (DCH) surrounded 
by a $1.5$-T solenoidal magnet, an electromagnetic calorimeter 
(EMC) comprising $6580$ CsI(Tl) crystals, and a detector of 
internally reflected Cherenkov light (DIRC) providing $K$--$\pi$ 
separation over the range of laboratory momentum relevant
for this analysis ($1.5$--$4.5\gevc$).

\section{ANALYSIS METHOD}
\label{sec:Analysis}
The analysis method is similar to that used in our previous measurement of these
quantities~\cite{BaBarSin2alpha2002}.  We reconstruct a sample of neutral $B$ mesons 
($B_{\rm rec}$) decaying to final states with two charged tracks, and
examine the remaining particles in each event to determine whether 
the second $B$ meson (\Btag) decayed as a $\Bz$ or $\Bzb$ (flavor tag).
The \CP\ asymmetry parameters in $\Bz\to\pip\pim$ decays are determined with a 
maximum likelihood fit including information about the flavor of \Btag\ and the
difference $\deltat$ between the decay times of the $B_{\rm rec}$ and \Btag\ decays.  
The decay rate distribution $f_+\,(f_-)$ when $B_{\rm rec}\to\pip\pim$ and 
$\Btag = \Bz\,(\Bzb)$ is given by
\begin{eqnarray}
\fpm = \frac{e^{-\left|\deltat\right|/\tau}}{4\tau} [1
& \pm & \spipi\sin(\deltamd\deltat) \nonumber \\
& \mp & \cpipi\cos(\deltamd\deltat)],
\label{fplusminus}
\end{eqnarray}
where $\tau$ is the mean $\Bz$ lifetime and $\deltamd$ is the mixing
frequency due to the neutral-$B$-meson eigenstate mass difference.

We first perform a maximum-likelihood fit
that uses kinematic, event-shape, and particle-identification information to
determine signal and background yields corresponding to the four 
distinguishable final states ($\pip\pim$, $\Kp\pim$, $\Km\pip$, $\Kp\Km$).  The
results of this fit are described in Ref.~\cite{BaBarAkpiPRL}, which reports
our observation of direct \CP\ violation in $\Bz\to\Kp\pim$ decays.  
The parameters $\spipi$ and $\cpipi$ are obtained from a second fit adding 
$B$-flavor and decay-time information, where the yields and $K\pi$
asymmetries for signal and backgrond events are fixed to the values obtained
in the first fit.

\subsection{Event Selection}
We reconstruct two-body neutral-$B$ decays from pairs of 
oppositely-charged tracks located within the geometric acceptance of the DIRC
and originating from a common decay point near the interaction region.
We require that each track have an associated Cherenkov-angle ($\theta_c$) 
measured with at least five signal photons detected in the DIRC, where the 
value of $\theta_c$ must agree within $4\sigma$ with either the pion or kaon particle 
hypothesis.  The last requirement efficiently removes events containing high-momentum 
protons.  Electrons are removed based on energy-loss measurements in the SVT and DCH, 
and on a comparison of the track momentum and associated energy deposited in the EMC.  

Identification of pions and kaons is primarily accomplished by including 
$\theta_c$ as a discriminating variable in the 
maximum likelihood fit.  We construct probability density functions (PDFs) 
for $\theta_c$ from a sample of approximately 
$430000$ $D^{*+}\to D^0\pi^+\,(\Dz\to\Km\pip)$ decays reconstructed in data, where 
$\Kmp/\pipm$ tracks are identified through the charge correlation with the
$\pipm$ from the $D^{*\pm}$ decay.  The PDFs are constructed separately for 
$\Kp$, $\Km$, $\pip$, and $\pim$ tracks as a function of momentum and polar angle 
using the measured and expected values of $\theta_c$, and its uncertainty.
The average $K$--$\pi$ separation, defined
as the difference between the expected angles for the kaon and pion mass
hypotheses divided by the average uncertainty, varies from $12$ 
standard deviations ($\sigma$) at a laboratory momentum of $1.5\gev/c$, to $2\sigma$
at $4.5\gevc$.  

Signal decays are identified using two kinematic variables: (1) the 
difference $\de$ between the energy of the $B$ candidate
in the $\epem$ center-of-mass (CM) frame and $\sqrt{s}/2$ and (2) the beam-energy 
substituted mass 
$\mes = \sqrt{(s/2 + {\mathbf {p}}_i\cdot {\mathbf {p}}_B)^2/E_i^2- {\mathbf {p}}_B^2}$.
Here, $\sqrt{s}$ is the total CM energy, and the $B$ momentum ${\mathbf {p_B}}$ 
and the four-momentum of the initial state $(E_i, {\mathbf {p_i}})$ are 
defined in the laboratory frame. 
For signal decays, $\de$ and $\mes$ are distributed according to Gaussian
distributions with resolutions of $27\mev$ and $2.6\mevcc$, respectively.  
The distribution of $\mes$ peaks near the $B$ mass for all four final states.  
To simplify the likelihood fit, we reconstruct the kinematics 
of the $B$ candidate using the pion mass for all tracks.  With this choice, 
$\Bz\to\pip\pim$ decays peak near $\de = 0$.  For $B$ decays with one or two 
kaons in the final state, the $\de$ peak position is shifted and parameterized 
as a function of the kaon momentum in the laboratory frame.  The average shifts 
with respect to zero are $-45\mev$ and $-91\mev$, respectively, and this 
separation in $\de$ provides additional discriminating power in the fit.  
We require $5.20 < \mes < 5.29\gevcc$ and 
$\left|\de\right|<150\mev$.  The large sideband region in $\mes$ is used to 
determine background-shape parameters, while the wide range in $\de$ allows us 
to separate $B$ decays to all four final states in the same fit.

We have studied potential backgrounds from higher-multiplicity $B$ decays 
and find them to be negligible near $\de = 0$.  The dominant source of background 
is the process $\epem\to q\bar{q}\; (q=u,d,s,c)$, which produces a distinctive 
jet-like topology.  In the CM frame we define the angle $\theta_S$ between the 
sphericity axis~\cite{sph} of the $B$ candidate and the sphericity axis of the 
remaining particles in the event.  For background events, $\left|\cos{\theta_S}\right|$ 
peaks sharply near unity, while it is nearly flat for signal decays.  We require 
$\left|\cos{\theta_S}\right|<0.8$, which removes approximately $80\%$ of this 
background.  Additional background suppression is accomplished by including
the Fisher discriminant ${\cal F}$ described in Ref.~\cite{BaBarSin2alpha2002}
as one of the variables in the maximum likelihood fit.

\subsection{\boldmath $B$-Flavor Identification and Decay-Time Reconstruction}
We use a multivariate technique~\cite{BaBarsin2beta} to determine the flavor of 
the $\Btag$ meson.  Separate neural networks are trained to identify primary leptons, 
kaons, soft pions from $D^*$ decays, and high-momentum charged particles from \B\ decays.  
Events are assigned to one of five mutually exclusive tagging categories 
based on the estimated average mistag probability and the source of the tagging information 
(Table~\ref{tab:tagging}).
The quality of tagging is expressed in terms of the effective efficiency 
$Q = \sum_k \epsilon_k (1-2w_k)^2$, where $\epsilon_k$ and $w_k$ are the 
efficiencies and mistag probabilities, respectively, for events tagged in category $k$.
Table~\ref{tab:tagging} summarizes the tagging performance measured in a data sample
\Bflav\ of fully reconstructed neutral $B$ decays to 
$D^{(*)-}(\pip,\, \rho^+,\, a_1^+)$.  The assumption of equal tagging efficiencies and 
mistag probabilities for signal $\pip\pim$, $\Kp\pim$, and $\Kp\Km$ decays is validated 
in a detailed Monte Carlo simulation.  Separate background efficiencies for the different 
decay modes are determined simultaneously with $\spipi$ and $\cpipi$ in the fit.

\begin{table}[!tbp]
\caption{Average tagging efficiency $\epsilon$, average mistag fraction $w$,
mistag fraction difference $\Delta w = w(\Bz) - w(\Bzb)$, and effective tagging efficiency 
$Q$ for signal events in each tagging category.  The quantities are measured in the 
\Bflav\ sample.}
\smallskip
\begin{center}
\begin{tabular}{crclrclrclrcl} \hline\hline
Category & \multicolumn{3}{c}{$\epsilon\,(\%)$} & \multicolumn{3}{c}{$w\,(\%)$} & \multicolumn{3}{c}{$\Delta w\,(\%)$} &
\multicolumn{3}{c}{$Q\,(\%)$} \rule[-2mm]{0mm}{6mm} \\\hline
{\tt Lepton}    & $9.6  $&$ \pm $&$ 0.1 $&$ 3.4  $&$ \pm $&$ 0.5 $&$ 0.1  $&$ \pm $&$ 0.8 $&$ 8.3  $&$ \pm $&$ 0.3$\\
{\tt Kaon\,I}   & $16.7 $&$ \pm $&$ 0.1 $&$ 8.6  $&$ \pm $&$ 0.5 $&$ -1.8 $&$ \pm $&$ 0.8 $&$ 11.4 $&$ \pm $&$ 0.3$\\
{\tt Kaon\,II}  & $19.3 $&$ \pm $&$ 0.1 $&$ 20.0 $&$ \pm $&$ 0.5 $&$ -3.0 $&$ \pm $&$ 0.9 $&$ 7.2  $&$ \pm $&$ 0.2$\\
{\tt Inclusive} & $20.1 $&$ \pm $&$ 0.1 $&$ 30.7 $&$ \pm $&$ 0.6 $&$ -4.4 $&$ \pm $&$ 0.9 $&$ 3.0  $&$ \pm $&$ 0.2$\\
{\tt Untagged}  & $34.3 $&$ \pm $&$ 0.2 $\\\hline
Total $Q$       &        &       &       &        &       &       &        &       &       &$ 29.9 $&$ \pm $&$ 0.5$ \rule[-2mm]{0mm}{6mm} \\\hline\hline
\end{tabular}
\end{center}
\label{tab:tagging}
\end{table}

The time difference $\deltat = \Delta z/\beta\gamma c$ is obtained from the 
known boost of the $\epem$ system ($\beta\gamma = 0.56$) and the measured distance 
$\Delta z$ along the beam ($z$) axis between the $B_{\rm rec}$ and $B_{\rm tag}$ decay
vertices.  We require $\left|\deltat\right|<20\ps$ and 
$\sigma_{\deltat} < 2.5\ps$, where $\sigma_{\deltat}$ is the error on $\deltat$ determined
separately for each event.  The resolution function for signal candidates is a sum of 
three Gaussians, identical to the one described in Ref.~\cite{BaBarsin2beta}, 
with parameters determined from a fit to the \Bflav\ sample (including events in all 
five tagging categories).  The background $\deltat$ distribution is modeled
as the sum of three Gaussian functions, where the common parameters used to describe the
background shape for all tagging categories are determined simultaneously with 
the \CP\ parameters in the maximum likelihood fit.

\subsection{Maximum Likelihood Fit}
We use an unbinned extended maximum likelihood fit to extract \CP\ parameters
from the $B_{\rm rec}$ sample.  The likelihood for candidate $j$ tagged in category 
$k$ is obtained by summing the product of event yield $n_{i}$, tagging efficiency 
$\epsilon_{i,k}$, and probability ${\cal P}_{i,k}$ over the eight possible signal 
and background hypotheses $i$ (referring to $\pi^{+}\pi^{-}$, $K^{+}\pi^{-}$, 
$K^{-}\pi^{+}$, and $K^{+}K^{-}$ combinations).  The extended likelihood function 
for category $k$ is
\begin{equation}
{\cal L}_k = \exp{\left(-\sum_{i}n_i\epsilon_{i,k}\right)}
\prod_{j}\left[\sum_{i}n_i\epsilon_{i,k}{\cal P}_{i,k}(\vec{x}_j;\vec{\alpha}_i)\right].
\end{equation}
The yields for the $K\pi$ final state are parameterized as 
$n_{\Kpm\pimp}=n_{K\pi}\left(1\mp \akpi\right)/2$, where $\akpi$
is the direct-\CP-violating asymmetry~\cite{BaBarAkpiPRL}.
The probabilities ${\cal P}_{i,k}$ are evaluated as the product of PDFs 
for each of the independent variables 
$\vec{x}_j = \left\{\mes, \de, {\cal F}, \theta_c^+, \theta_c^-, \deltat\right\}$, 
where $\theta_c^+$ and $\theta_c^-$ are the Cherenkov angles for the positively- and 
negatively-charged tracks.  The $\deltat$ PDF for signal $\pip\pim$ decays is given 
by Eq.~\ref{fplusminus}, modified to include the $w_k$ and $\Delta w_k$ for each tag 
category, and convolved with the signal resolution function.  The $\deltat$ PDF for 
signal $K\pi$ decays takes into account $\Bz$--$\Bzb$ mixing and the correlation 
between the charge of the kaon and the flavor of $\B_{\rm tag}$.

There are $46$ free parameters in the fit:
\begin{itemize}
 \item $12$ parameters describing the background PDFs for $\mes$, $\de$, and
 $\fish$;
 \item $8$ parameters describing the background $\deltat$ PDF;
 \item $12$ background flavor-tagging efficiencies;
 \item $12$ background flavor-tagging efficiency asymmetries; and
 \item $\spipi$ and $\cpipi$.
\end{itemize}
The signal and background yields and $K\pi$ asymmetries were determined in
a separate fit that does not use flavor-tagging or $\deltat$ information~\cite{BaBarAkpiPRL}.
Out of a fitted sample of $68030$ events, we find $n_{\pi\pi} = 467\pm 33$, 
$n_{K\pi} = 1606\pm 51$, and $n_{KK}=3\pm 12$ decays, and measure $\akpi = -0.133\pm 0.030$, 
where all errors are statistical only.  We fix $\tau$ and $\deltamd$ to their world-average 
values~\cite{PDG2004}.  The total likelihood ${\cal L}$ is the product of likelihoods for 
each tagging category, and the free parameters are determined by maximizing the quantity 
$\ln{\cal L}$.

\section{PHYSICS RESULTS}
\label{sec:Physics}

The fit to the $B_{\rm rec}$ sample yields
\begin{eqnarray*}
\spipi & =          & -0.30\pm 0.17\,({\rm stat})\pm 0.03\,({\rm syst}),\\
\cpipi & =          & -0.09\pm 0.15\,({\rm stat})\pm 0.04\,({\rm syst}).
\end{eqnarray*}
The correlation between $\spipi$ and $\cpipi$ is $-1.6\%$, and the
correlations with all other free parameters are less than $1\%$.  These 
preliminary results are consistent with our previously published 
measurements~\cite{BaBarSin2alpha2002}, and are combined with our measurements
of the branching fractions for the isospin-related decay modes $\pipm\piz$ and
$\piz\piz$ to determine model-independent bounds on $\alpha$ (see Ref.~\cite{isospin}).

\begin{figure}[!tbp]
\begin{center}
\includegraphics[width=8cm]{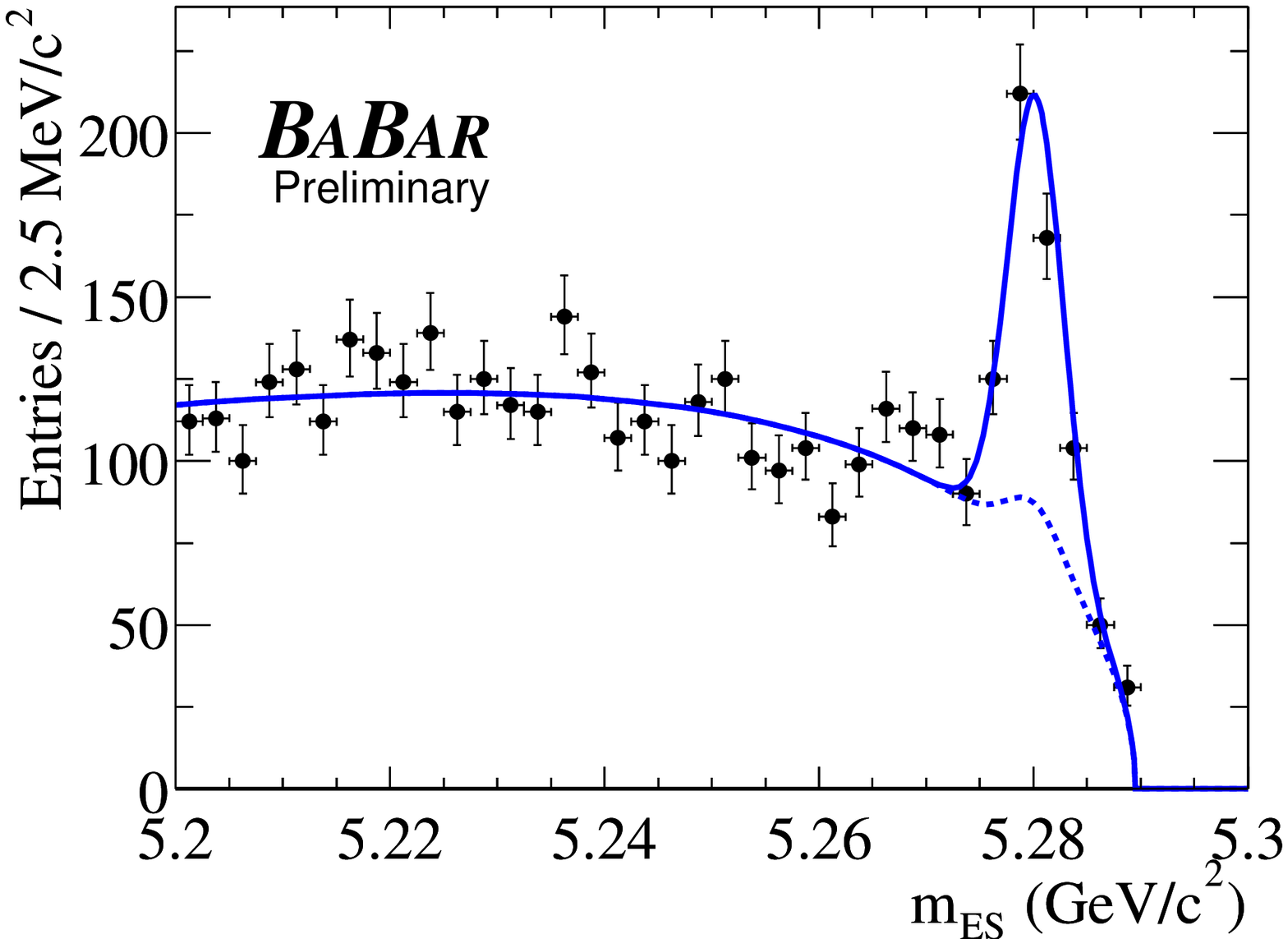}
\includegraphics[width=8cm]{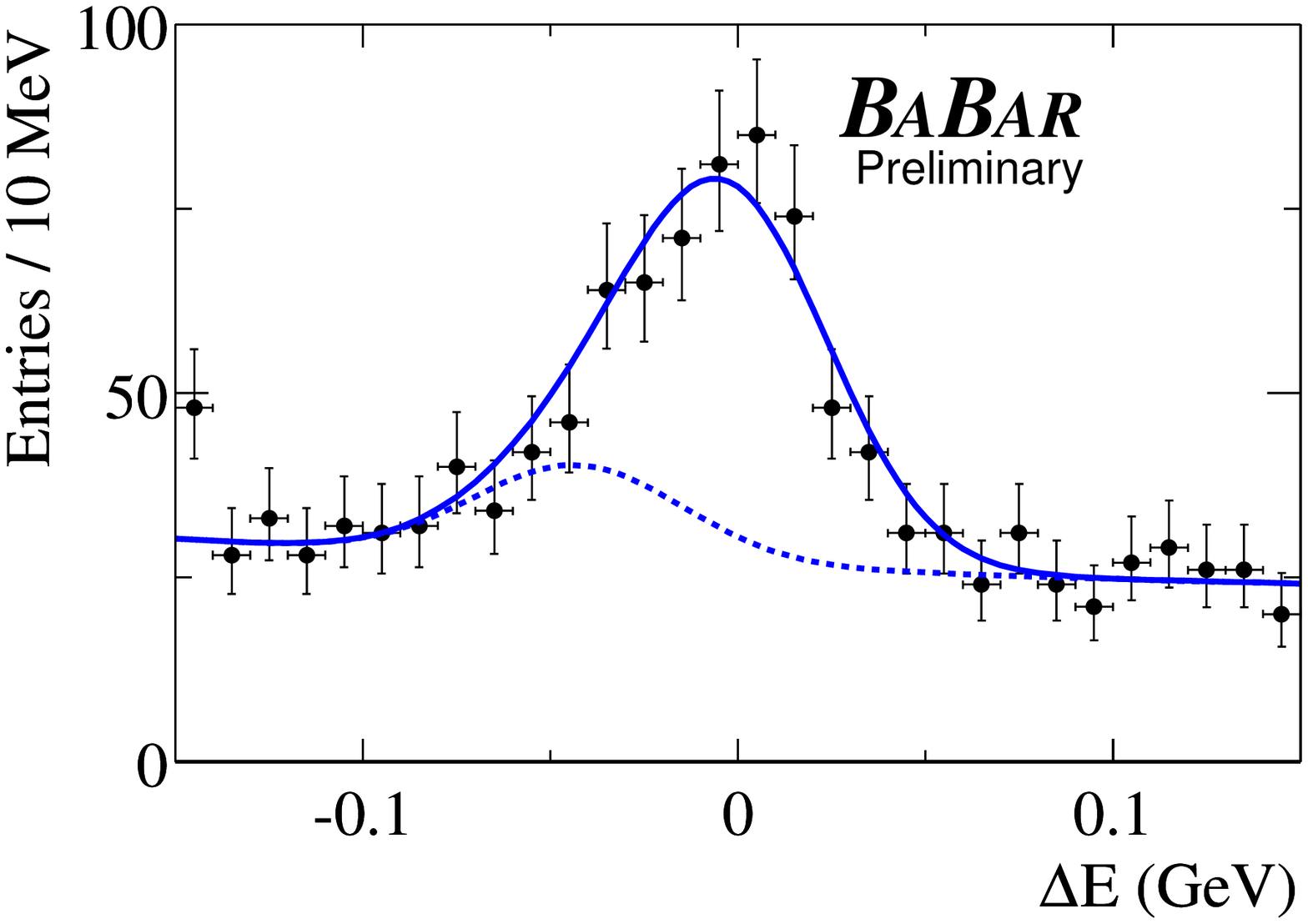}
\caption{Distributions of (left) $\mes$ and (right) $\de$ for
events (points with error bars) enhanced in signal $\pip\pim$ decays using
additional requirements on probability ratios.  
Solid curves represent projections of the maximum likelihood fit, dashed
curves represent $q\bar{q}$ and $K\pi\to\pi\pi$ cross-feed
background.}
\label{fig:pipi}
\end{center}
\end{figure}

\begin{figure}[!tbp]
\begin{center}
\includegraphics[height=10cm]{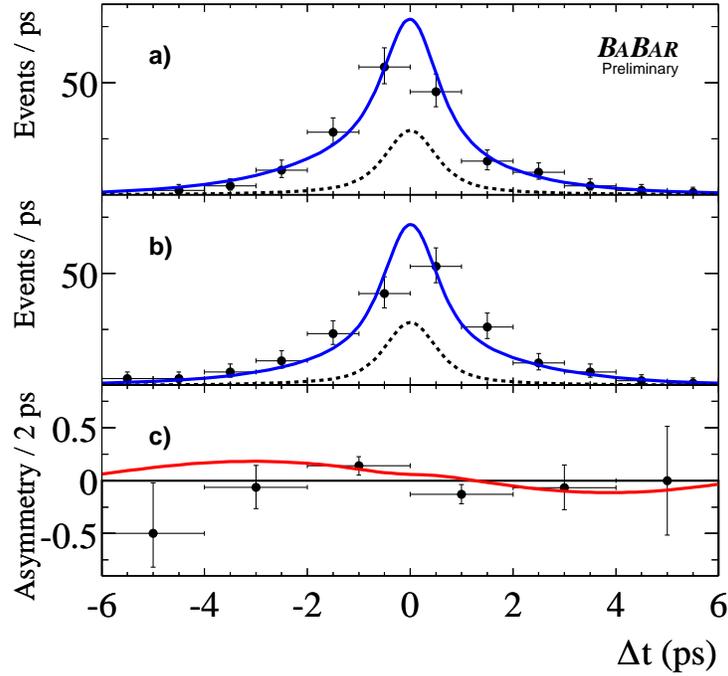}
\caption{Distribution of the decay-time difference $\deltat$ for events enhanced in 
signal $\Bz\to\pip\pim$ decays using additional requirements on 
probability ratios.  The top two plots show events where $B_{\rm tag}$ is
identified as (a) $\Bz$ ($n_{\Bz}$) or (b) $\Bzb$ ($n_{\Bzb}$).  
Solid curves indicate the projection of the maximum likelihood fit including signal 
and background, while the dashed curves show the contribution from background events.  
(c) The asymmetry (points with errors), defined as 
$\left(n_{\Bz} - n_{\Bzb}\right)/\left(n_{\Bz} + n_{\Bzb}\right)$, for different
bins in $\deltat$, and the projection of the full fit (solid curve).}
\label{fig:asym}
\end{center}
\end{figure}

Figure~\ref{fig:pipi} shows distributions of $\mes$ and $\de$ for events enhanced
in signal $\pip\pim$ decays.  We apply additional requirements on probability ratios
based on all PDFs except the variable being plotted.  The solid curves are projections 
of the maximum likelihood fit for the sum of signal and all background components, 
while the dashed curves indicate the sum of $q\bar{q}$ and $\Bz\to\Kp\pim$ cross-feed 
background.  Figure~\ref{fig:asym} shows distributions of $\deltat$ for events with 
$B_{\rm tag}$ tagged as $\Bz$ or $\Bzb$, and the asymmetry as a function of $\deltat$ 
for signal $\pi\pi$ decays selected with requirements on probability ratios including 
all PDFs except $\deltat$.  

As a consistency check on the $\deltat$ resolution function, we 
perform a $\Bz$--$\Bzb$ mixing study using the large number of $K\pi$
signal decays in the $B_{\rm rec}$ sample.  Floating $\tau$ and $\deltamd$ along 
with $\spipi$, $\cpipi$, and $\akpi$, we find values consistent with the world averages 
($\tau = 1.60\pm 0.04\,{\rm ps}$ and $\deltamd = 0.523\pm 0.028\,{\rm ps}^{-1}$), 
and the CP parameters are consistent
with the nominal fit results.

\section{SYSTEMATIC STUDIES}
\label{sec:Systematics}
Table~\ref{tab:sys} summarizes the contributions to the total systematic uncertainty
from the dominant sources.  These include imperfect knowledge of the PDF shape 
parameters; the $B$-flavor-tagging parameters (Table~\ref{tab:tagging}); the alignment
of the SVT; the event-by-event beam-spot position; 
the potential effect of doubly Cabibbo-suppressed decays of the $B_{\rm tag}$ 
meson~\cite{Owen}, and the $B$ lifetime and mixing frequency.  In addition, to 
confirm that we are sensitive to non-zero values of $\spipi$ and $\cpipi$, we fit a large 
sample of Monte-Carlo simulated signal decays with large values of the \CP\ parameters.  
The fit results are consistent with the generated values, and we assign the 
sum in quadrature of the statistical uncertainty and the difference between the fitted 
and generated values as a conservative systematic error accounting for potential bias.  
The effect of uncertainty on the signal and background yields and 
$K\pi$ asymmetries is negligible for both $\spipi$ and $\cpipi$.  The total systematic 
uncertainty is calculated by summing in quadrature the individual contributions.

\begin{table}[!htb]
\caption{Summary of systematic uncertainties on $\spipi$ and $\cpipi$ (see text
for details).  The total uncertainty is calculated as the sum in quadrature of the 
individual contributions.}
\begin{center}
\begin{tabular}{ccc} \hline\hline
Source            	    & $\spipi$ & $\cpipi$ \\
\hline
PDF parameters              & $ 0.017$    & $ 0.018$ \\
$B$-flavor identification   & $ 0.005$    & $ 0.015$ \\
SVT alignment               & $ 0.010$    & $ 0.002$ \\
Beam spot                   & $ 0.010$    & $ 0.010$ \\
Tag-side interference       & $ 0.008$    & $ 0.023$ \\
$\tau_{\Bz}$ and $\deltamd$ & $ 0.001$    & $ 0.004$ \\
Potential bias              & $ 0.013$    & $ 0.007$ \\
\hline
Total                       & $ 0.027$    & $ 0.035$ \\
\hline\hline
\end{tabular}
\end{center}
\label{tab:sys}
\end{table}

\section{SUMMARY}
\label{sec:Summary}
In summary, we present preliminary updated measurements of the \CP-violating asymmetries
$\spipi$ and $\cpipi$ occuring in the time distributions of $\Bz\to\pip\pim$ decays.  We
find $\spipi = -0.30\pm 0.17\pm 0.03$ and $\cpipi = -0.09\pm 0.15\pm 0.04$, which are
consistent with our previous measurements and with the hypothesis $\spipi = \cpipi = 0$.
These results do not confirm the observation of large \CP\ violation reported in 
Ref.~\cite{BelleSin2alpha2004}.

\section{ACKNOWLEDGMENTS}
\label{sec:Acknowledgments}

% Standard acknowledgments paragraph; must always be included.
We are grateful for the 
extraordinary contributions of our \pep2\ colleagues in
achieving the excellent luminosity and machine conditions
that have made this work possible.
The success of this project also relies critically on the 
expertise and dedication of the computing organizations that 
support \babar.
The collaborating institutions wish to thank 
SLAC for its support and the kind hospitality extended to them. 
This work is supported by the
US Department of Energy
and National Science Foundation, the
Natural Sciences and Engineering Research Council (Canada),
Institute of High Energy Physics (China), the
Commissariat \`a l'Energie Atomique and
Institut National de Physique Nucl\'eaire et de Physique des Particules
(France), the
Bundesministerium f\"ur Bildung und Forschung and
Deutsche Forschungsgemeinschaft
(Germany), the
Istituto Nazionale di Fisica Nucleare (Italy),
the Foundation for Fundamental Research on Matter (The Netherlands),
the Research Council of Norway, the
Ministry of Science and Technology of the Russian Federation, and the
Particle Physics and Astronomy Research Council (United Kingdom). 
Individuals have received support from 
CONACyT (Mexico),
the A. P. Sloan Foundation, 
the Research Corporation,
and the Alexander von Humboldt Foundation.

\end{document}